\documentclass[aip, numerical, amsmath, amssymb, jcp]{revtex4-1}

\usepackage{graphicx}
\usepackage{dcolumn}
\usepackage{bm}

\usepackage[utf8]{inputenc}
\usepackage[T1]{fontenc}
\usepackage{mathptmx}
\usepackage{color}

\begin{document}

\title{Charge transfer via deep hole in the J51/N2200 blend}

\author{Xiaoyu Xie}
\email{njuxxy2013@gmail.com}
\affiliation{School of Chemistry and Chemical Engineering, Nanjing University, Nanjing 210023, China}
\author{Chunfeng Zhang}
\affiliation{National Laboratory of Solid State Microstructures, School of Physics, and Collaborative Innovation Center of Advanced Microstructures, Nanjing University, Nanjing 210093, China.}
\author{Haibo Ma}
\email{haibo@nju.edu.cn}
\affiliation{School of Chemistry and Chemical Engineering, Nanjing University, Nanjing 210023, China}

\date{\today}

\begin{abstract}
In recently developed non-fullerene acceptor (NFA) based organic solar cells (OSCs), both the donor and acceptor parts can be excited by absorbing light photons. Therefore, both electron transfer and hole transfer channels could occur at the donor/acceptor interface for generating free charge carriers in NFA based OSCs. However, in many molecular and DNA systems, recent studies revealed the high charge transfer (CT) efficiency cannot be reasonably explained by a CT model with only highest occupied molecular orbitals (HOMOs) and lowest unoccupied molecular orbitals (LUMOs) of donor and acceptor molecules. In this work, taking an example of a full-polymer blend consisting of benzodithiophenealt-benzotriazole copolymers (J51) as donor and naphthalene diimide-bithiophene (N2200) as acceptor, in which the ultrafast hole transfer has been recently reported, we investigate its CT process and examine the different roles of various frontier molecular orbitals. Through a joint study of quantum mechanics electronic structure calculation and nonadiabatic dynamics simulation, we find the hole transfer between HOMOs of J51 and N2200 can hardly happen but the hole transfer from HOMO of N2200 to HOMO-1 of J51 is much more efficient. This points out the underlying importance of deep hole channel in CT process and indicates that including frontier molecular orbitals (FMOs) other than HOMOs and LUMOs is highly necessary to build a robust physical model for studying CT process in molecular optoelectronic materials.
\end{abstract}

\maketitle

\section{Introduction} \label{sec: 1}
Charge transfer (CT) process is one of the key fundamental steps in many molecular optoelectronic materials, ranging from organic solar cells (OSCs) \cite{deibel2010role, clarke2010charge, lee2010charge, gelinas2014ultrafast, deotare2015nanoscale, yao2016ultrafast, kohn2017impact, pelzer2017molecular, fusella2018band, huang2019charge} to DNA-based nanoelectronics \cite{genereux2010mechanisms, kawai2013hole, xiang2015intermediate, renaud2016deep}. In organic systems, the interaction between the photo-generated electron-hole pair is very strong with typical binding energy magnitudes of 0.3–1 eV, in large excess of $k_\mathrm{B}T$, due to the weakly Coulomb screening with a relative low dielectric permittivity (around 2-3), which is quite distinct from their inorganic counterparts. \cite{bernardi2016computer} Thus, a donor/acceptor (D/A) interface utilizing energy offsets between frontier molecular orbitals (FMOs) in electron donor and acceptor moieties, is usually adopted in these materials to drive an efficient CT before the photoexcitation recombines either radiatively or non-radiatively. Depending on the energetic driving force comes from the photoexcitation in electron donor or acceptor moiety, the CT process is usually classified to electron or hole transfer, i.e. transferring an electron from the lowest unoccupied molecular orbital (LUMO) of donor to the LUMO of acceptor or transferring a hole from the highest occupied molecular orbital (HOMO) of electron acceptor to the HOMO of electron donor. In conventional OSCs, the electron transfer is considered to dominate the CT at D/A interface \cite{clarke2010charge, lee2010charge, grancini2013hot, lane2015hot}, but hole transfer was recently addressed to be crucial in new OSCs with new non-fullerene acceptor molecules \cite{kim2015flexible, wang2019all, lee2019recent, liu202018}, in which the power conversion efficiency (PCE) is boosted to 18.2 \%. \cite{liu202018} 

Traditionally, energetic and coupling information of only HOMOs and LUMOs is used in building simplified models for the qualitative or semi-quantitative interpretation of CT processes in molecular systems. In 2014, one of the authors (HM) and Troisi suggested the possibility of additionally utilizing a hot electron via LUMO+1 or a deep hole via HOMO-1 to enhance the CT efficiency by more than two orders of magnitude through reducing the energy barrier for CT, based on an investigation of 5 electron donor molecules and 6 acceptor ones.\cite{ma2014modulating} This hypothesis was later supported by a statistical analysis of experimental data from 80 high performing non-fullerene electron acceptors \cite{kuzmich2017trends} and used in building quantitative structure-property relationship (QSPR) models for predicting PCEs of new OSC molecules \cite{sahu2018toward, sahu2019designing, sahu2019unraveling}. In 2015, the generation of hot charge carriers was experimentally verified by terahertz photoconductivity measurements by Lane, \emph{et al.} \cite{lane2015hot}, in which they found the transfer of an electron following excitation of ZnPc at 400 or 615nm to a higher unoccupied MO of C$_{60}$. Through combining a density functional theory (DFT) calculation and fewest-switches surface-hopping (FSSH) dynamics simulations of the heterojunction of pentacene and monolayer MoS$_2$, Cui, \emph{et al.} \cite{xie2019photoinduced} also found the hole at MoS$_2$ transfers to HOMO-1 of pentacene at the beginning steps after photoexcitation at monolayer MoS$_2$. In 2016, Renaud, \emph{et al.} reported a similar deep hole CT process in DNA hairpins, with an enhancement in CT rates by one hundredfold and much weaker distance dependence. \cite{renaud2016deep} In their work, the authors adopted two different electron acceptors, stilbenedicarboxamide (SA) and naphthalene-diimide (NDI) at the tail of a DNA hairpin with a same donor stilbenediether (SD) at the head of this DNA hairpin and compared their charge migration performances. Due to the difference in electronic structures of the two acceptors, a much faster and more distance-insensitive CT process has been found in NDI-DNA-SD system via deep hole transfer. All of these studies indicate that, deep occupied MOs (e.g. HOMO-1, HOMO-2 of donor) and high unoccupied MOs (LUMO+1, LUMO+2 of acceptor) other than HOMOs and LUMOs are also highly necessary to be incorporated in a quantitative CT model and this provides a new promising way to enhance CT efficiency in molecular materials by manipulating their energetics.

\begin{figure}
    \includegraphics[scale=0.5]{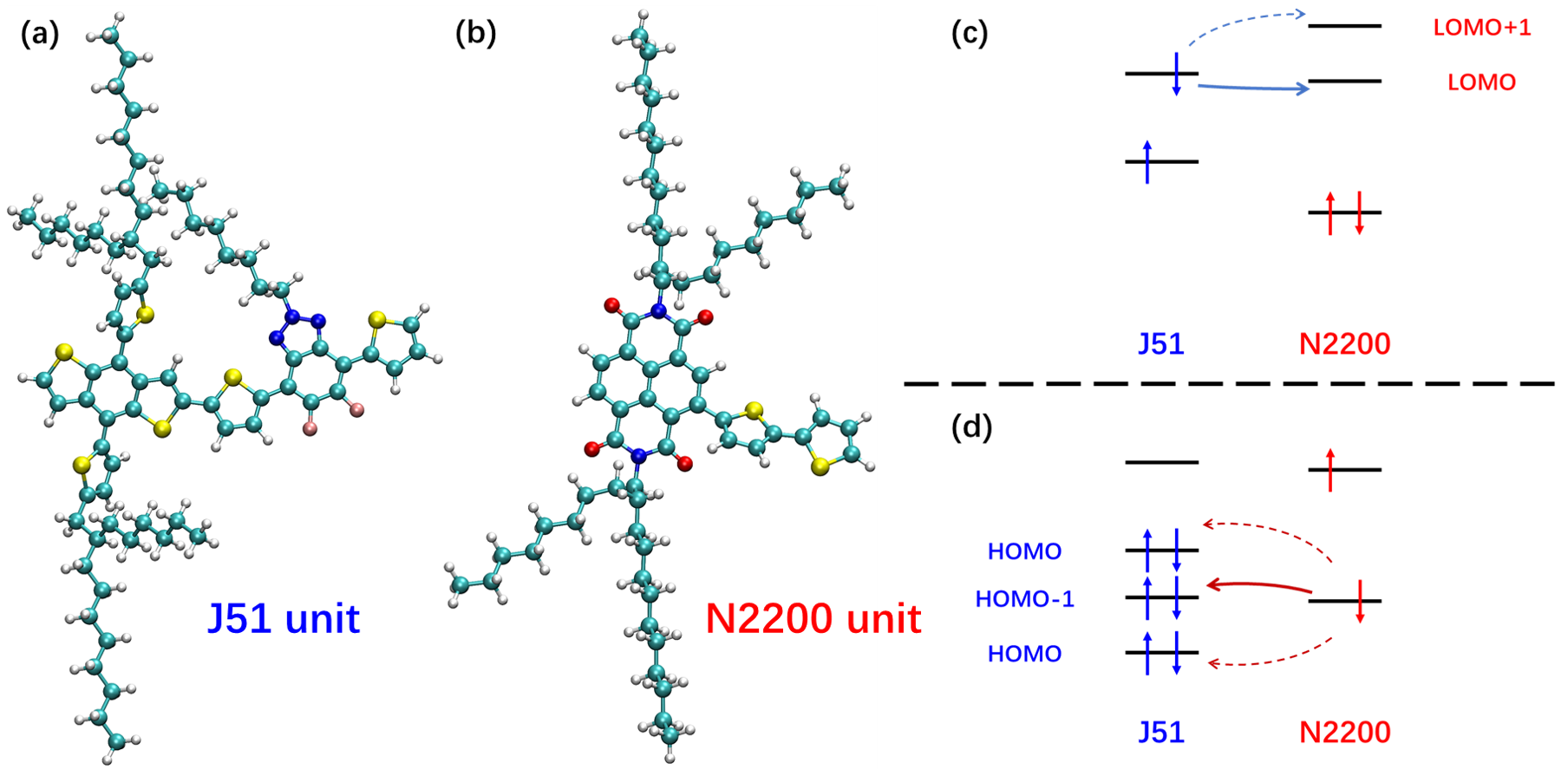}
    \caption{(a) Unit structure of J51 (donor), (b) unit structure of N2200 (acceptor), (c) schematic electron transfer and (d) schematic hole transfer via deep hole approach for charge transfer process in J51/N2200 blend.} \label{fig: 1}
\end{figure}

Recently, by utilizing ultrafast optical measurements, we found ultrafast hole transfer mediated by intra-moiety polaron pairs \cite{wang2019ultrafast} in the OSC system (PCE: 8.27\%) \cite{gao2016all} consisting of donor benzodithiophenealt-benzotriazole copolymers (J51, whose unit is shown in Fig.~\ref{fig: 1} (a)) and acceptor naphthalene diimide-bithiophene (N2200, whose unit is shown in Fig.~\ref{fig: 1} (b)). In this work, we further explore the detailed CT mechanism between J51 and N2200 polymers and investigate the possibility of a deep hole channel by calculating the electronic structure and simulating non-adiabatic dynamics behaviours in J51/N2200 blend. We find that hole transfer between HOMOs of J51 and N2200 can hardly happen but hole transfer from HOMO of N2200 to HOMO-1 of J51 is much more efficient with a ultrafast time scale around the magnitude of a few picoseconds, qualitatively agreeing with our previous experimental measurement. This verifies again the underlying importance of deep hole in CT process and indicates that including FMOs other than HOMOs and LUMOs is highly necessary to build a robust physical model for studying CT process in molecular optoelectronic materials.

\section{Methodology and computational details} \label{sec: 2}
\subsection{Electron-vibration interaction model} \label{subsec: 2.1}
Electron-vibration (el-vib) interaction model is a popular and powerful tool to study exciton/charge processes in molecular aggregated systems. By adopting a linear approximation (i.e. truncating high order el-vib interaction), the Hamiltonian for an el-vib coupled system can be expressed as,
\begin{equation} \label{eq: Hamiltonian}
    \hat{H}=\hat{H}_{el}+\hat{H}_{vib}+\hat{H}_{el-vib},
\end{equation}
with
\begin{equation} \label{eq: Hel}
    \hat{H}_{el}=\sum_{i}\varepsilon_{i}\vert\psi_i\rangle\langle\psi_i\vert+\sum_{i\neq j}V_{ij}\vert\psi_i\rangle\langle\psi_j\vert,
\end{equation}
\begin{equation} \label{eq: Hvib}
    \hat{H}_{vib}=\sum_{I}\frac{1}{2}\hbar\omega_{I}(-\frac{\partial^2}{\partial Q_I^2}+Q_I^2),
\end{equation}
\begin{equation} \label{eq: Helvib}
    \hat{H}_{el-vib}=\sum_{i,I}g^{I}_{i}Q_I\vert\psi_i\rangle\langle\psi_i\vert + \sum_{i\neq j,I}g^{I}_{ij}Q_I\vert\psi_i\rangle\langle\psi_j\vert.
\end{equation}
Here, $\varepsilon_{i}$ and $V_{ij}$ represent the energy of the electronic state $\vert\psi_i\rangle$ and the electronic coupling between $\vert\psi_i\rangle$ and $\vert\psi_j\rangle$ under the equilibrium geometry, respectively. $\omega_I$ is the frequency of the vibration mode $I$ with $Q_I$ being its dimensionless displacement. In the el-vib coupling term, $g_{i}^I$ are local el-vib couplings and $g_{ij}^I$ are non-local ones.

Considering there are numerous vibration modes in the condensed phase system, spectral density is alternatively used to describe the strength of el-vib couplings with respect to frequency of modes,
\begin{equation} \label{eq: spectral density}
    J_{i/ij}(\omega) = \frac{1}{2\hbar}\sum_I{g^I_{i/ij}}^2\delta(\omega-\omega_I).
\end{equation}

\subsection{Computational details of parameters in the model} \label{subsec: 2.2}
\begin{figure}
    \includegraphics[scale=0.5]{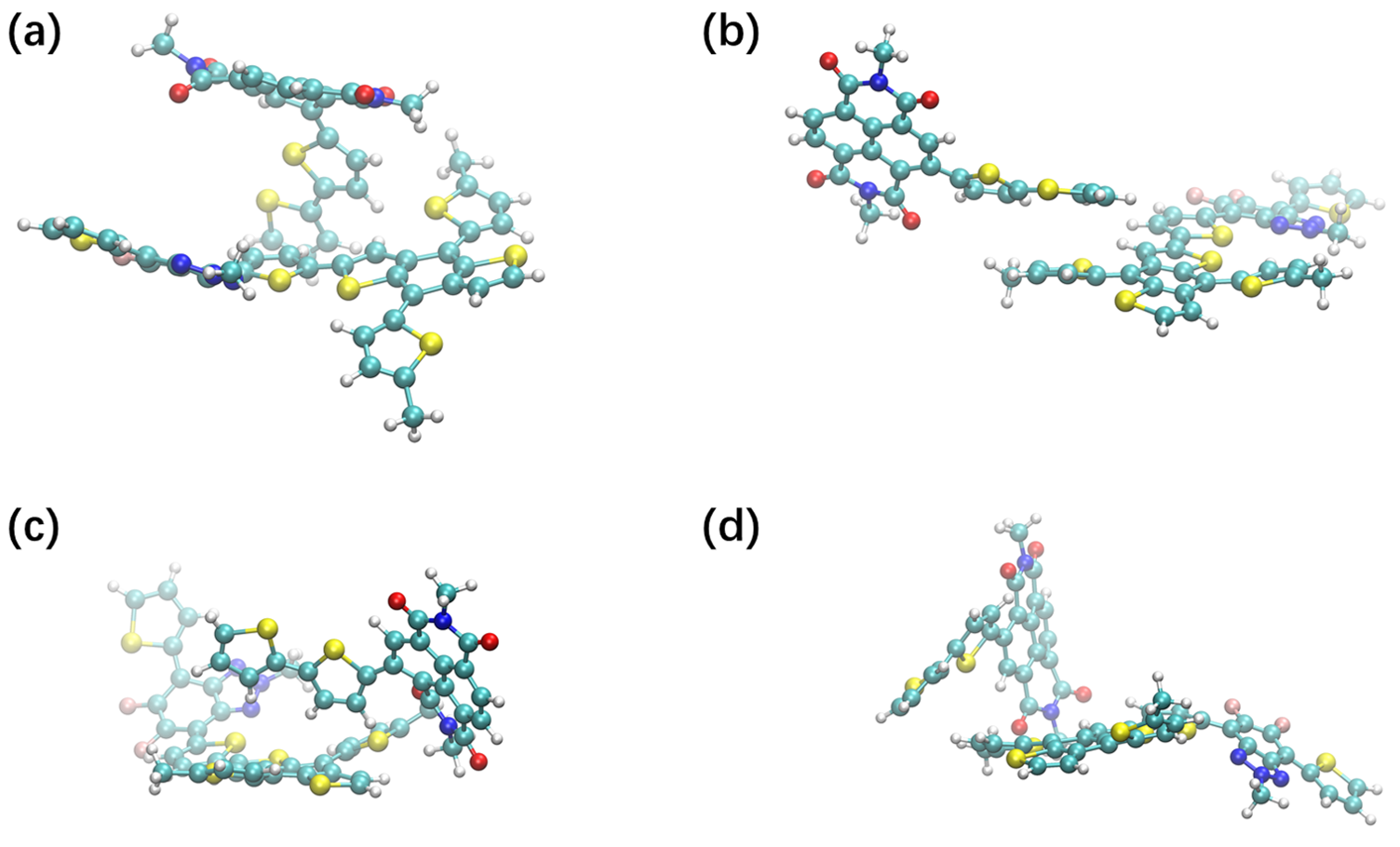}
    \caption{Selected donor-acceptor pairs for electronic structure calculation (only conjugation skeletons are used): (a) pair 1, (b) pair 2, (c) pair 3 and (d) pair 4.} \label{fig: 2}
\end{figure}

Since the system (polymeric J51/N2200 blend) in our work is amorphous and quite large for \textit{ab initio} calculation, we combine classical molecular dynamics (MD) simulation of an amorphous cell of polymer J51/N2200 blend and quantum mechanical (QM) calculation of neighboring pairs of N2200 and J51 units (see Fig.~\ref{fig: 2}): 1). For $\varepsilon_{i}$ and $V_{ij}$ in Eq.~\ref{eq: Hel}, we adopt the average values of electronic parameters from sampled MD conformations; 2). the Fourier transform of their time autocorrelation function provides information of electron-vibration interaction (i.e. spectral density in Eq.~\ref{eq: spectral density}).

For the MD part, we build an amorphous cell with periodic boundary condition containing 4 chains of polymer N2200 and J51, with each polymer of J51 and N2200 consisting of 5 units as shown in Fig.~\ref{fig: 1} (a) and (b) respectively. Then, MD simulation is performed with COMPASSII force field \cite{sun2016compass}, in which NPT ensemble (300 K and 1 atm) is used for both geometry relaxation (1 fs per time step until density and temperature of system get nearly equilibrium, which are $\rho\sim 1.024\pm0.004$ g $\cdot$ cm$^{-3}$ and $T\sim 297.9\pm 2.3$ K in our case) and dynamics simulation (10 ps with 1 fs per time step). All MD simulations are done using Materials Studio package \cite{module2013material}.  In order to illustrate the feasibility of deep hole transfer in J51/N2200 blend, four nearest-neighboring pairs of N2200 and J51 unit (alkyl chains are simplified to reduce computational costs, see Fig.~\ref{fig: 2}) are selected after the MD simulation, and their trajectories are sampling for the subsequent electronic structure calculation and further analysis.

Considering the weak binding energy of intra-moiety polaron pairs found in this system \cite{wang2019ultrafast}, we take the one-electron approximation to describe of charge transfer at the donor/acceptor interface, which is widely adopted in theoretical studies of charge transfer in molecular systems \cite{webster1991solvation, stuchebrukhov2003tunneling, shkrob2007structure}, i.e. the electronic states $\vert\psi_i\rangle$ in Eq.~\ref{eq: Hel} would be reduced to FMOs $\vert\varphi_i\rangle$. A correlation analysis for the energies of anionic states by DFT and time-dependent DFT (TDDFT) in supplementary Fig.~S1 verifies the validation of one-electron approximation in our system. Consequently, parameters in electronic Hamiltonian (Eq.~\ref{eq: Hel}) are approximated as the Fock matrix elements of the donor-acceptor system.
\begin{eqnarray} \label{eq: block diagonal}
    \varepsilon_i(t) &=& \pm\langle\varphi_i(t)\vert\hat{F}(t)\vert\varphi_i(t)\rangle, \\
    V_{ij}(t) &=& \langle\varphi_i(t)\vert\hat{F}(t)\vert\varphi_j(t)\rangle,
\end{eqnarray}
with $\hat{F}$ being the Fock operator of system and $t$ representing the specific conformation of MD trajectory at time $t$. The sign of $\varepsilon_i(t)$ depends on if it is an occupied MO: '$-$' for  occupied MO (hole transfer) and '$+$' for unoccupied MO (electron transfer). Then, we can obtain parameters of electronic Hamiltonian by averaging $\varepsilon_i(t)$ and $V_{ij}(t)$ through the time. In practice, we perform the block-diagonal approach \cite{domcke1991theory, kondov2007quantum, xie2015full} to construct localised FMOs and calculate $\varepsilon_i(t)$ and $V_{ij}(t)$. DFT calculations are applied to compute the electronic structure (i.e. Fock matrix) of the pair systems at M062X\cite{zhao2008the}/6-311g(d, p) level by using Gaussian16 package \cite{g16}. The sampling time interval $\Delta t$ of conformations is 5 fs with total time $t_\mathrm{total}=$ 10 ps. i.e. 2000 snapshots are used for each pair in Fig.~\ref{fig: 2}.

Time-dependent value ($\varepsilon_i(t)$ and $V_{ij}(t)$) also contains information of el-vib interaction \cite{may2011charge, aghtar2012juxtaposing},
\begin{equation} \label{eq: spectral density via MD}
    J_{i/ij}(\omega)=\frac{2}{\pi\hbar}\mathrm{tanh}\left(\frac{\hbar\omega}{2k_\mathrm{B}T}\right)\int_0^\infty\mathrm{d}tC_{i/ij}(t)\mathrm{cos}(\omega t),
\end{equation}
here $k_\mathrm{B}T$ is the thermal energy with $T=$ 300 K in this case and $C_{i/ij}(t)$ is the autocorrelation function of time-dependent $\varepsilon_i(t)/V_{ij}(t)$,
\begin{eqnarray}
    C_i(t) &=& \langle\varepsilon_i(t)\varepsilon_i(0)\rangle - {\varepsilon_i}^2, \label{eq: Ci} \\
    C_{ij}(t) &=& \langle V_{ij}(t)V_{ij}(0)\rangle - {V_{ij}}^2. \label{eq: Cij}
\end{eqnarray}
To reduce the effect of truncated integral problem in Eq.~\ref{eq: spectral density via MD}, a cosine function $g(t)=\cos{(\pi t/t_0)}$ with $t_0=t_\mathrm{total}$ is multiplied to autocorrelation function to make the spectral density smooth. In our case, we only consider local spectral density terms for the further dynamics simulation. The non-local terms would be treated in other way, which will be discussed in next Section (Section~\ref{subsec: 2.3}).

\subsection{The Redfield dynamics simulation} \label{subsec: 2.3}
Since the degrees of freedom (DoF) of nuclear motion are too large to be treated accurately by a full quantum dynamics theory in a discrete way, a common solution is using time-dependent equations of reduced density matrix $\rho(t)$ (i.e. quantum master equations) by separating total system into system ($\hat{H}_{el}$) and bath terms ($\hat{H}_{vib}$). In our case, we apply the Redfield equation \cite{bloch1957generalized, redfield1965ibm, may2011charge, breuer2002theory, berkelbach2013microscopicI, tempelaar2018vibronicIII} for the system with only local el-vib interaction. in the basis of eigenstates of $\hat{H}_{el}$ ($\hat{H}_{el}\vert\alpha\rangle=\varepsilon_\alpha\vert\alpha\rangle$), the Redfield equation is expressed as,
\begin{equation} \label{eq: Redfield}
    i\hbar\frac{\partial}{\partial t}\rho_{\alpha\beta}(t)=(\varepsilon_\alpha-\varepsilon_\beta)\rho_{\alpha\beta}(t)-i/\hbar\sum_{\gamma\delta}R_{\alpha\beta\gamma\delta}(t)\rho_{\gamma\delta}(t),
\end{equation}
where $R$ is the Redfield tensor and given by,
\begin{equation} \label{eq: Redfield tensor}
    R_{\alpha\beta\gamma\delta} = \Gamma_{\delta\beta\alpha\gamma}^++\Gamma_{\delta\beta\alpha\gamma}^--\delta_{\delta\beta}\sum_{\kappa}\Gamma_{\alpha\kappa\kappa\gamma}^+-\delta_{\alpha\gamma}\sum_{\kappa}\Gamma_{\delta\kappa\kappa\beta}^-,
\end{equation}
with
\begin{eqnarray}
    \Gamma_{\alpha\beta\gamma\delta}^+(t) &=& \int_0^t\mathrm{d}\tau e^{-i(\varepsilon_\gamma-\varepsilon_\delta)\tau/\hbar}\sum_iC_i^{\mathrm{ther}}(\tau)\langle\alpha\vert\varphi_i\rangle\langle\varphi_i\vert\beta\rangle\langle\gamma\vert\varphi_i\rangle\langle\varphi_i\vert\delta\rangle, \label{eq: Gamma plus} \\
    \Gamma_{\alpha\beta\gamma\delta}^-(t) &=& \int_0^t\mathrm{d}\tau e^{-i(\varepsilon_\alpha-\varepsilon_\beta)\tau/\hbar}\sum_i{C_i^{\mathrm{ther}}}^*(\tau)\langle\alpha\vert\varphi_i\rangle\langle\varphi_i\vert\beta\rangle\langle\gamma\vert\varphi_i\rangle\langle\varphi_i\vert\delta\rangle. \label{eq: Gamma minus}
\end{eqnarray}
Here, $C_i^{\mathrm{ther}}(t)$ is the thermal autocorrelation function of $\varepsilon_i$,
\begin{equation} \label{eq: thermal C}
    C_i^{\mathrm{ther}}(t) = \int_0^\infty\mathrm{d}\omega J_i(\omega)\left(\coth{\frac{\hbar\omega}{2k_\mathrm{B}T}\cos{\omega t} - i \sin{\omega t}}\right),
\end{equation}
where $T=300$ K is applied in this case. In practice, the Markovian approximation ($R_{\alpha\beta\gamma\delta}(t)=R_{\alpha\beta\gamma\delta}(+\infty)$) and the secular approximation ($R_{\alpha\beta\gamma\delta} = 0 $ for $\varepsilon_\alpha-\varepsilon_\beta\neq\varepsilon_\gamma-\varepsilon_\delta$) are used \cite{berkelbach2013microscopicI}.

As shown in Eq.~\ref{eq: Gamma plus}, \ref{eq: Gamma minus} and \ref{eq: thermal C}, we only consider the influence of local el-vib couplings. For the non-local terms, we combine the Redfield equation with the frozen-modes (FM) approach \cite{montoya2015extending}. Since the non-local el-vib couplings are predominated by slow intermolecular modes \cite{xie2018nonlocal}, the fluctuation of electronic couplings which caused by nuclear motion can be regarded as static disorders. Thus, the FM approach could be applied to take the non-local el-vib couplings into account. In our work, we replace $\hat{H}_{el}$ by a new $\hat{H}_{el}^\prime$ with,
\begin{eqnarray}
    \langle\varphi_i\vert\hat{H}_{el}^\prime\vert\varphi_i\rangle &=& \varepsilon_i, \label{eq: FM diagonal} \\
    \langle\varphi_i\vert\hat{H}_{el}^\prime\vert\varphi_j\rangle &=& V_{ij} + \Delta V_{ij}\,\,\mathrm{with}\,\,i\neq j. \label{eq: FM off-diagonal}
\end{eqnarray}
Here $\Delta V_{ij}$ is a Gaussian-type random number, whose standard deviation $\sigma_{ij}$ can be obtained from the time-dependent $V_{ij}(t)$. Then, the results of dynamics simulation is computed as statistical average of 1000 trajectory samplings.

\section{Results and discussion} \label{sec: 3}
\subsection{Electronic structures of MD trajectory snapshots} \label{subsec: 3.1}
\begin{figure}
    \includegraphics[scale=0.45]{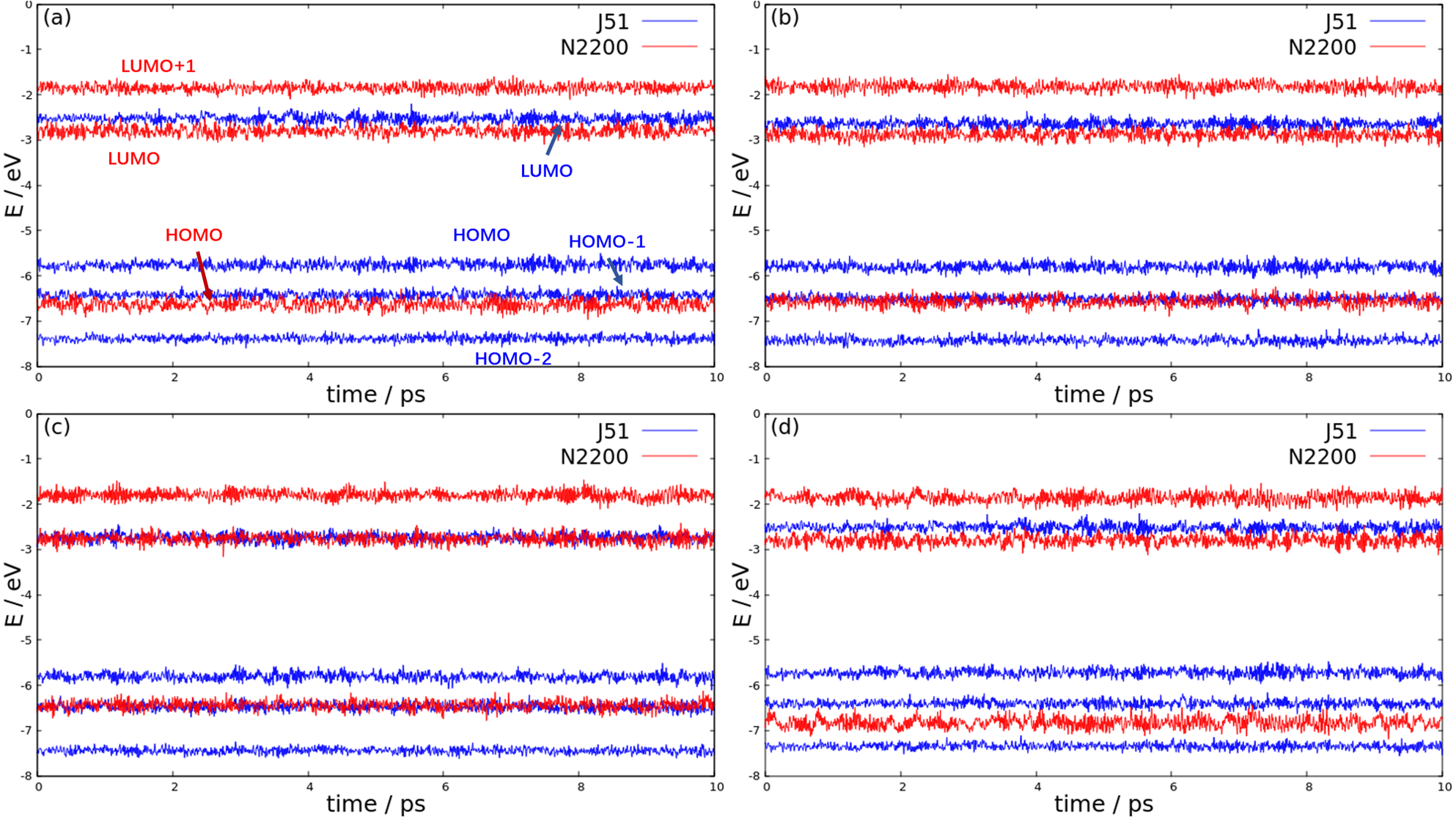}
    \caption{The energies of localised frontier molecular orbitals (FMOs) of (a) pair 1, (b) pair 2, (c) pair 3 and (d) pair 4. Here blue and red for donor and acceptor respectively.} \label{fig: 3}
\end{figure}

By using the block diagonal method, we calculate the orbital energies of localised FMOs and electronic couplings between these orbitals for the four neighboring donor-acceptor pairs. The results of orbital energies are shown in Fig.~\ref{fig: 3}, we also list the average values in Tab.~\ref{tab: 1}. The energy orders of orbitals for these four pairs are similar and consistent with our schemes in Fig.~\ref{fig: 1} (c) and (d). For the electron transfer channel, the energies of LUMO+1 of acceptor are significantly higher than LUMO of donor. Therefore, only LUMO-LUMO electron transfer is possible in this system. While there are two occupied FMOs of donor whose orbital energies are higher than the orbital energy of HOMO of acceptor or close to it, which implies that there might be more than one path for the hole transfer channel. So, in this work we will focus only on the hole transfer channel in polymeric J51/N2200 blend. Through examining the spatial locations of FMOs of J51 and N2200 units, we find that HOMO and LUMO of N2200 unit locate locally at bithiophene part and naphthalene diimide part respectively, which is consistent with the idea of intra-moiety polaron-pairs.

\begin{table}
    \caption{The average energies of localised FMOs for four pairs with superscript representing donor (D) or acceptor (A) molecules. (Unit: eV)} \label{tab: 1}
    \begin{ruledtabular}
    \begin{tabular}{cccccccc}
               & HOMO-2$^\mathrm{D}$ & HOMO-1$^\mathrm{D}$ & HOMO$^\mathrm{D}$ & HOMO$^\mathrm{A}$ & LUMO$^\mathrm{A}$ & LUMO+1$^\mathrm{A}$ & LUMO$^\mathrm{D}$ \\
        \hline
        Pair 1 &               -7.39 &               -6.43 &             -5.76 &             -6.63 &             -2.80 &               -1.85 &             -2.53 \\
        Pair 2 &               -7.43 &               -6.51 &             -5.80 &             -6.57 &             -2.88 &               -1.82 &             -2.64 \\
        Pair 3 &               -7.44 &               -6.48 &             -5.81 &             -6.44 &             -2.77 &               -1.80 &             -2.74 \\
        Pair 4 &               -7.35 &               -6.41 &             -5.72 &             -6.83 &             -2.80 &               -1.86 &             -2.51 \\
    \end{tabular}
    \end{ruledtabular}
\end{table}

\begin{figure}
    \includegraphics[scale=0.28]{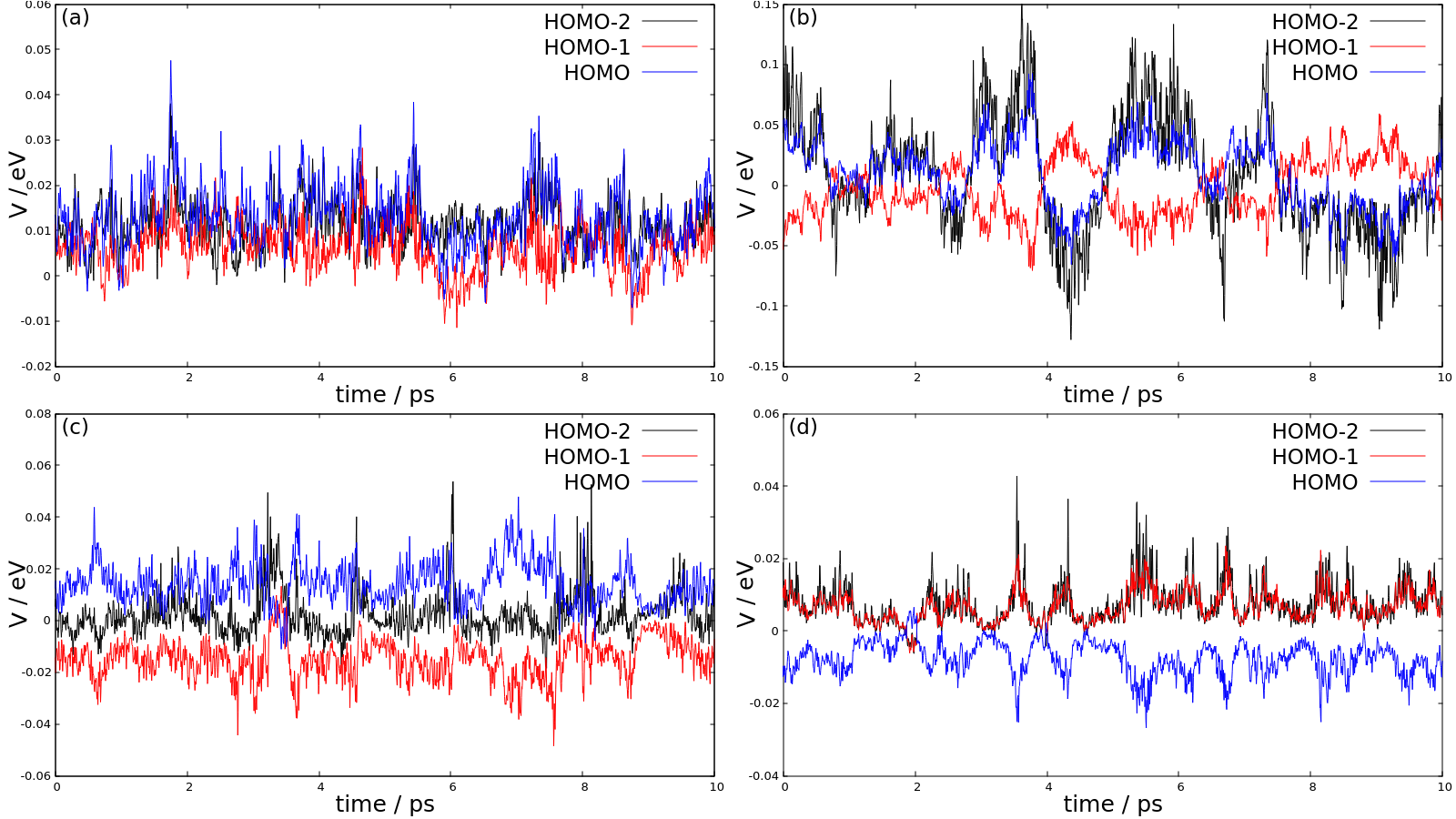}
    \caption{The electronic couplings between HOMO of acceptor and occupied FMOs of donor for (a) pair 1, (b) pair 2, (c) pair 3 and (d) pair 4.} \label{fig: 4}
\end{figure}

In Fig.~\ref{fig: 4}, we present results of the electronic couplings $V(t)$ between HOMO of acceptor and three occupied FMOs of donor. The three couplings behave very similar trajectories in all four pairs (pair 1, pair 2, pair 4 and HOMO-1/HOMO of pair 3) through time, which may result from the similar $\pi$-orbital structures of occupied FMOs of the $\pi$-conjugated-structure donor. Unlike the trajectories of energy part in Fig.~\ref{fig: 3}, there are long-wave oscillations for electronic couplings in Fig.~\ref{fig: 4} (also see the non-local spectral densities in supplementary Fig.~S2), which implies the strong interaction between low frequency modes and electronic couplings. The slow oscillations again validate the FM sampling. The average values and their standard deviations are listed in Tab.~\ref{tab: 2}, where the values of these couplings and the standard deviation are around 5-15 meV and 5-10 meV respectively, except the large fluctuations in pair 2.

\begin{table}
    \caption{Average electronic couplings (standard deviation) between HOMO of acceptor and occupied FMOs of donor in four pairs. (Unit: meV)} \label{tab: 2}
    \begin{ruledtabular}
    \begin{tabular}{cccc}
               &  HOMO-2$^\mathrm{D}$ &   HOMO-1$^\mathrm{D}$ &    HOMO$^\mathrm{D}$ \\
        \hline
        Pair 1 &        11.53  (5.18) &          6.09  (5.22) &        13.00  (6.91) \\
        Pair 2 &         8.25 (47.38) &         -3.13 (22.41) &         8.79 (27.25) \\
        Pair 3 &         2.25  (8.27) &        -14.86  (7.56) &        14.02  (7.95) \\
        Pair 4 &         7.32  (5.44) &          6.65  (4.10) &        -7.58  (4.68) \\
    \end{tabular}
    \end{ruledtabular}
\end{table}

Both data in Tab.~\ref{tab: 1} and \ref{tab: 2} are used to construct electronic Hamiltonian ($\hat{H}_{el}$ in Eq.~\ref{eq: Hel}, $\hat{H}_{el}^\prime$ in Eq.~\ref{eq: FM diagonal} and \ref{eq: FM off-diagonal}) for the dynamics simulations of hole transfer in the next Section (Section~\ref{subsec: 3.2}).

\subsection{Dynamics results of the Markovian Redfield simulation} \label{subsec: 3.2}
\begin{figure}
    \includegraphics[scale=0.28]{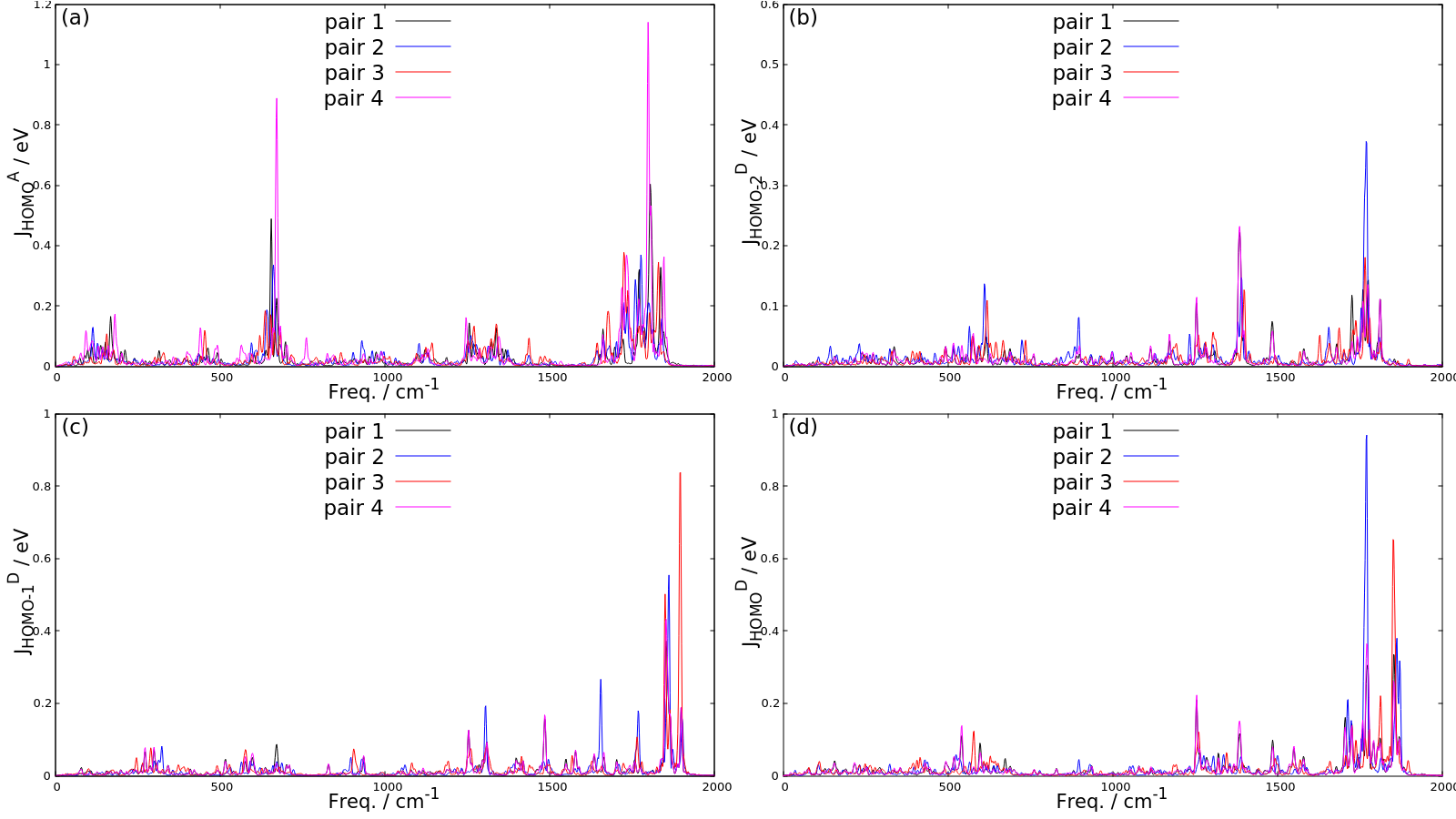}
    \caption{The spectral densities of the local el-vib couplings for (a) HOMO of acceptor, (b) HOMO-2 of donor, (c) HOMO-1 of donor and (d) HOMO of donor.} \label{fig: 5}
\end{figure}

To perform the Redfield and Redfield-FM simulation for the hole transfer, information of local el-vib couplings is needed. We compute spectral densities the four orbitals of four pairs via Eq~\ref{eq: spectral density via MD}. The results are illustrated in Fig.~\ref{fig: 5}, and four pairs show similar peaks for all four orbitals. In this figure, we truncate frequency at 2000 cm$^{-1}$ because the signal is very weak for modes with $\omega>$ 2000 cm$^{-1}$. The reason of the weak signal is that the modes in this region are C-H stretch modes and they do not affect the conjugated skeletons much. Below 2000 cm$^{-1}$ ($\sim$ 0.25 eV), a general feature of local spectral density is that high frequency modes ($\sim$ 1700-1800 cm$^{-1}$) are dominant. In this region, vibration modes are C-C stretch of the conjugated core and they could change the $\pi$-structure of FMOs and affect their energies. For HOMO of acceptor N2200 unit, there are non-neglected peaks at around 600 cm$^{-1}$, the modes in this region can be assigned as breath modes of conjugated core, which affect the structure and energies of FMOs of N2200 unit, too. Furthermore, we could evaluate the reorganization energy $\lambda_i$ of each orbital $i$ using $\lambda_i=\int\mathrm{d}\omega J_i(\omega)\omega$, and the results are listed in Tab.~\ref{tab: 3}.

\begin{table}
    \caption{The reorganization energise of of localised FMOs for four D/A pairs. (Unit: meV)} \label{tab: 3}
    \begin{ruledtabular}
    \begin{tabular}{ccccc}
               & HOMO-2$^\mathrm{D}$ & HOMO-1$^\mathrm{D}$ & HOMO$^\mathrm{D}$ & HOMO$^\mathrm{A}$ \\
        \hline
        Pair 1 &               33.40 &               44.96 &             58.63 &             98.38 \\
        Pair 2 &               45.90 &               40.71 &             56.23 &             97.26 \\
        Pair 3 &               34.46 &               51.01 &             60.63 &             88.57 \\
        Pair 4 &               35.30 &               44.63 &             57.93 &            140.30 \\
    \end{tabular}
    \end{ruledtabular}
\end{table}

\begin{figure}
    \includegraphics[scale=0.28]{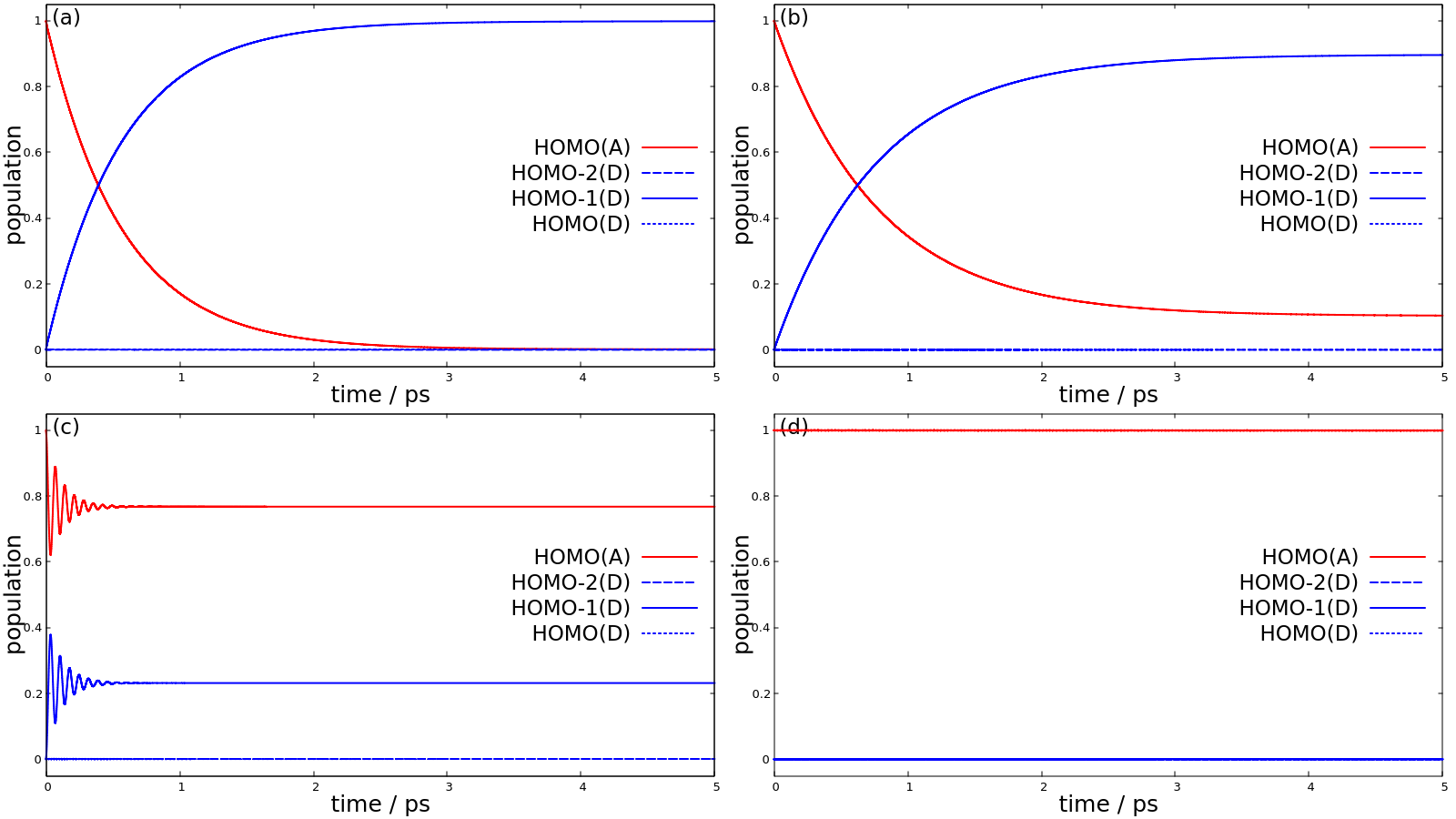}
    \caption{The Redfield simulation results without FM approach for (a) pair 1, (b) pair 2, (c) pair 3 and (d) pair 4.} \label{fig: 6}
\end{figure}

With the information in Tab.~\ref{tab: 1}, Tab.~\ref{tab: 2} (average coupling result) and Fig.~\ref{fig: 5}, the Markovian-approximated Redfield theory is applied to simulate the hole transfer processes for these four pairs with the initial hole localised at the HOMO of acceptor. The simulation results are shown in Fig.~\ref{fig: 6}, which are different from each other. First, the energy gap between HOMO of acceptor and HOMO/HOMO-2 of donor are too large for transfer of hole population. Deep hole transfers via HOMO-1 of donor occur for pair 1, 2 and 3 due to the small energy gap (0.20, 0.06 and -0.04 eV respectively) between HOMO of acceptor and HOMO-1 of donor. With the help of nuclear motion, hole population transfers from HOMO of acceptor to HOMO-1 of donor.  In pair 1 and 2, this transfer is exothermal and the population signal of HOMO-1 of donor appears around picoseconds. Besides, the larger energy gap and electronic coupling in pair 1 lead to a more complete and faster transfer. While the endothermal hole transfer in pair 3 converge very fast after the delay oscillation (decoherence). In pair 4, the gap between HOMO-1 of donor and HOMO of acceptor (0.42 eV) is too large to undergo the deep hole transfer, according to our Markovian-approximated Redfield simulation.

Furthermore, we consider the fluctuation of electronic couplings caused by vibration modes using the Redfield-FM approach. In our work, independent Gaussian disorders \cite{bassler1993charge} are added to three non-zero electronic couplings based on the value of standard deviation (Tab.~\ref{tab: 2}), which is contradictory with the fact that the dynamical disorders of these couplings are highly correlated (see Fig.~\ref{fig: 4} and discussion in Section~\ref{subsec: 3.1}). But the main dynamics feature of our systems is like typical quasi-two-level-systems (see Fig.~\ref{fig: 6}), which means only one of electronic coupling dominate the CT dynamics. Accordingly, our simplification of independent modeling of the electronic coupling fluctuations would be still valid.

\begin{figure}
    \includegraphics[scale=0.28]{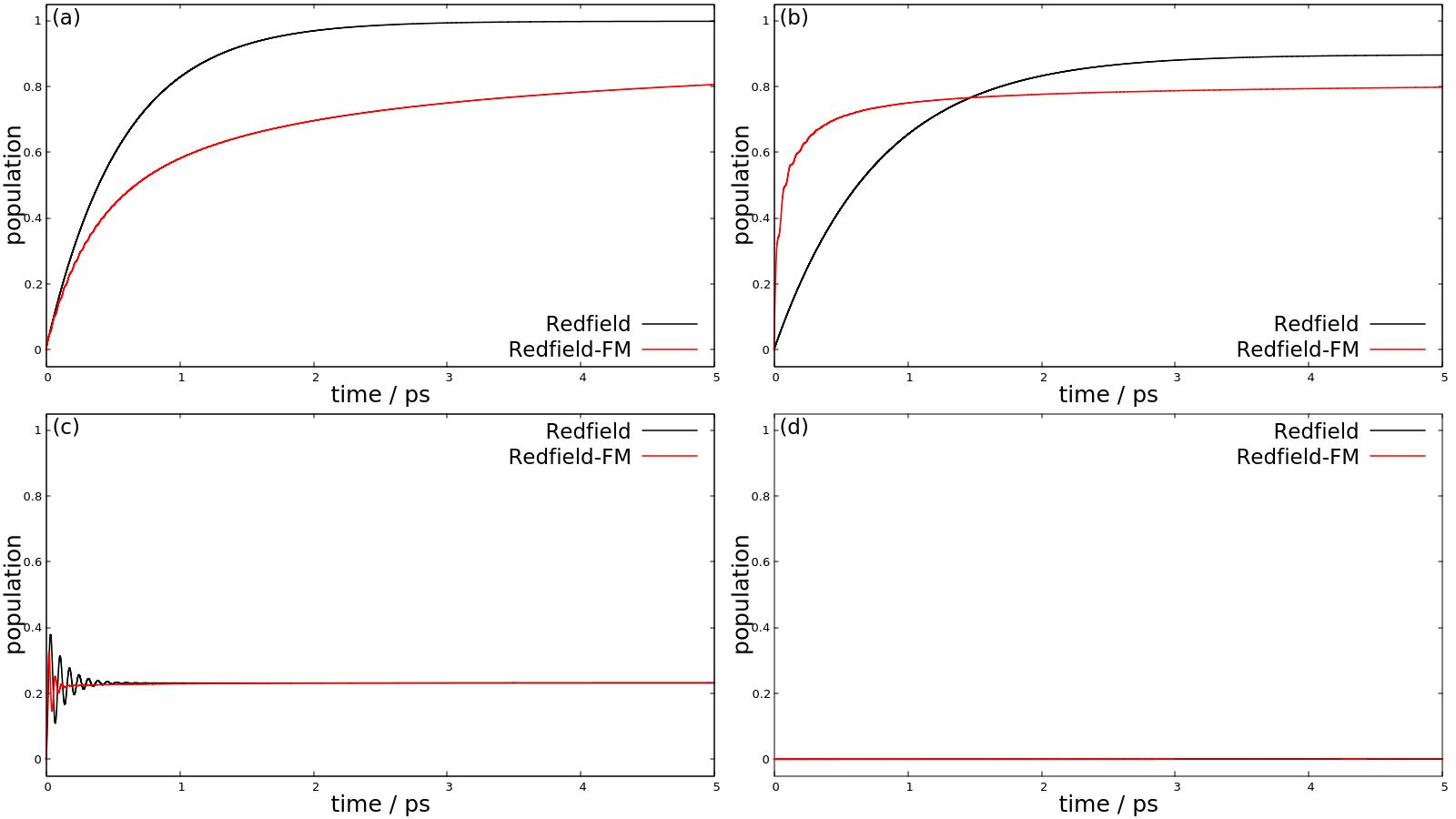}
    \caption{The comparison of time-evolution hole population located at HOMO-1 of donor between the Redfield equation and the Redfield-FM method for (a) pair 1, (b) pair 2, (c) pair 3 and (d) pair 4. }\label{fig: 7}
\end{figure}

The time-evolution of hole population at HOMO-1 of donor are presented in Fig.~\ref{fig: 7}, comparing to the results of the Redfield equation without FM approach. For exothermal transfer (pair 1 and 2), the final populations at HOMO-1 of donor both decrease, but the changing behaviours of rate are different. When using FM-Redfield, the rate decreases for pair 1 ($V>\sigma$) while the rate is increased for pair 2 ($V<\sigma$). Similar effect has been observed previously in simulating charge mobility of organic semiconductors. i.e. fluctuation of transfer integral slows down mobility of charge in the region of small non-local el-vib coupling \cite{atroisi2006charge}, while it enhances charge mobility when this coupling is strong enough to undergo the hopping-like charge diffusion \cite{song2015a, lian2019non}. For the endothermal transfer process (Fig.\ref{fig: 7} (c)), the fluctuation of electronic coupling is found to be beneficial for the decoherence between two states. In systems with large energy gap between donor and acceptor orbitals, the fluctuation still can not promote the efficient hole transfer (see Fig.\ref{fig: 7} (d)). However, this is actually an artifact of Markovian Redfield due to the used secular approximation. Our comparative simulation by non-Markovian Redfield and full quantum time-dependent density matrix renormalization group (tDMRG) found it's still feasible to have population signal of HOMO-1 of donor for pair 4 (see supplementary Fig.~S3(d)).

Therefore, our Redfield results can be used for only qualitative evaluation and we should notice our systems don't fulfill one of the pre-conditions ($V > \lambda$) of the quantitative success of Redfield theory. Our benchmark comparisons (see supplementary Fig.~S3) with non-Markovian Redfield and tDMRG simulations for the model in Fig.~\ref{fig: 6} indicate that there are noticeable deviations of our Redfield simulations from more accurate full quantum tDMRG results. But the comparisons in supplementary Fig.~S3 also show that our main conclusion of the possibility of deep hole transfer still holds on for all methods.

\section{Summary} \label{sec: 4}
In this work, we investigate the participation of different FMOs in the ultrafast CT process at J51/N2200 interface. By combining MD simulation of the amorphous blend and QM electronic structure calculation of four selected near-neighboring pairs, we build an el-vib interaction model and simulate nonadiabatic dynamics of hole transfer process of these pairs. Electronic structure calculation shows that energies of HOMO and HOMO-1 of J51 unit are higher than HOMO of N2200, suggesting a high possibility of deep hole transfer in J51/N2200 blend. Time-dependent orbital energies and further spectral density results demonstrate that local el-vib couplings are predominated by high frequency intramolecular modes, while low frequency nuclear motions play important roles for non-local el-vib couplings. With the el-vib model, we simulate the hole transfer processes using the Markovian Redfield equation, the results show that deep hole transfer from HOMO-1 of J51 unit to HOMO of N2200 unit is the only effective path, whose time scale is qualitatively close to the experimental finding (3 ps). This indicates that the deep occupied MOs (e.g. HOMO-1, HOMO-2 of donor) and high unoccupied MOs (e.g. LUMO+1, LUMO+2 of acceptor) other than HOMOs and LUMOs are also highly necessary to be incorporated in a quantitative CT model as well. It also provides a new promising way to enhance CT efficiency in molecular materials by manipulating their energetics.

\section{Supplementary materials}
The supplementary materials consist of a support information (SI) file and a zip file. The SI file includes the detailed methods of electronic structure calculation, validation of one-electron approximation, the results of non-local spectral densities and additional dynamics results via non-Markovian and tDMRG methods. The zip file contains xyz files of four trajectories for the QM calculation.

\section{Acknowledgments}
The work was supported by the National Natural Science Foundation of China (Grant No. 21673109) and the Fundamental Research Funds for the Central Universities (0204-14380126).

\section{Data availability}
The data that support the findings of this study are available from the corresponding author upon reasonable request.

\bibliography{ref}

\end{document}


\title{Supporting Information for "Charge transfer via deep hole in the J51/N2200 blend"}

\author{Xiaoyu Xie}
\email{njuxxy2013@gmail.com}
\affiliation{School of Chemistry and Chemical Engineering, Nanjing University, Nanjing 210023, China}
\author{Chunfeng Zhang}
\affiliation{National Laboratory of Solid State Microstructures, School of Physics, and Collaborative Innovation Center of Advanced Microstructures, Nanjing University, Nanjing 210093, China.}
\author{Haibo Ma}
\email{haibo@nju.edu.cn}
\affiliation{School of Chemistry and Chemical Engineering, Nanjing University, Nanjing 210023, China}

\date{\today}

\maketitle

\section{Electronic structure calculation}
Using one-electron approximation, the electronic Hamiltonian matrix of electronic states can be simplified as the Fock matrix of molecular orbitals (MOs). i.e.
\begin{equation}
\langle\varphi_1\varphi_2\ldots\varphi_i\vert\hat{H}_{el}\vert\varphi_1\varphi_2\ldots\varphi_j\rangle=\langle\varphi_i\vert\hat{F}\vert\varphi_j\rangle.
\end{equation}

Generally, MOs would be expanded as linear combination of a basis set (e.g. atomic orbitals $\{\phi\}$). Therefore,
\begin{equation}
\langle\varphi_i\vert\hat{F}\vert\varphi_j\rangle = \sum_{a,b}c_{ia}^*c_{jb}\mathbf{F}_{ab},
\end{equation}
here $c_{ia}$ is MO coefficient under an atomic orbital basis set ($\vert\varphi_i\rangle = \sum_{a}c_{ia}\vert\phi_a\rangle$).

In our case, we perform a block diagonalisation of Fock matrix \cite{domcke1991theory, kondov2007quantum, xie2015full} to construct local orbitals and calculate Fock matrix with respect to this orbital basis. We list detailed steps below for a system of donor/acceptor pair (D/A).

\begin{enumerate}
\item Calculating the Fock matrix $\textbf{F}_0$ with an atomic orbitals (AOs) basis set \{$\vert\phi^\mathrm{D}\rangle$,$\vert\phi^\mathrm{A}\rangle$\} for the full system.
\item Using overlap matrix $\textbf{S}$ to orthogonalise the basis set and obtain the Fock matrix with the orthogonalised basis set \{$\vert\phi_n^\prime\rangle$\}.
\begin{equation}
\vert\phi_n^\prime\rangle=\sum_i(\textbf{S}^{-\frac{1}{2}})_{in}\vert\phi_i\rangle,
\end{equation}
and
\begin{equation}
\textbf{F}=\textbf{S}^{-\frac{1}{2}}\textbf{F}_0\textbf{S}^{-\frac{1}{2}}.
\end{equation}
\item Assuming that step 2 does not break the localisation feature of the basis set, we rewrite Fock matrix $\textbf{F}$ into a block structure
\begin{equation}
\textbf{F}=\left(
\begin{array}{cc}
\textbf{F}_\mathrm{AA} & \textbf{F}_\mathrm{DA} \\
\textbf{F}_\mathrm{AD} & \textbf{F}_\mathrm{DD}
\end{array}\right),
\end{equation}
and block diagonalize the Fock matrix using eigenvectors of $\textbf{F}_\mathrm{AA}$ ($\textbf{C}_\mathrm{AA}$) and $\textbf{F}_\mathrm{DD}$ ($\textbf{C}_\mathrm{DD}$), i.e.
\begin{equation}
\overline{\textbf{F}}=\left(
\begin{array}{cc}
\textbf{C}_\mathrm{AA} & \textbf{0} \\
\textbf{0} & \textbf{C}_\mathrm{DD}
\end{array}\right)^{\dagger}\left(
\begin{array}{cc}
\textbf{F}_\mathrm{AA} & \textbf{F}_\mathrm{DA} \\
\textbf{F}_\mathrm{AD} & \textbf{F}_\mathrm{DD}
\end{array}\right)\left(
\begin{array}{cc}
\textbf{C}_\mathrm{AA} & \textbf{0} \\
\textbf{0} & \textbf{C}_\mathrm{DD}
\end{array}\right).
\end{equation}
\end{enumerate}

By using this approach and averaging electronic Hamiltonian of snapshots from MD, we could establish 4 times 4 electronic Hamiltonian for all four pairs. The results of Hamiltonian elements are listed below.

\begin{table}[ht]
    \caption{The electronic Hamiltonian elements for hole transfer of pair 1. (Unit: eV)}
    \begin{ruledtabular}
    \begin{tabular}{ccccc}
                            & HOMO$^\mathrm{A}$ & HOMO-2$^\mathrm{D}$ & HOMO-1$^\mathrm{D}$ & HOMO$^\mathrm{D}$ \\
        \hline
        HOMO$^\mathrm{A}$   &           6.63420 &             0.01153 &             0.00609 &           0.01300 \\
        HOMO-2$^\mathrm{D}$ &           0.01153 &             7.38608 &             0.00000 &           0.00000 \\
        HOMO-1$^\mathrm{D}$ &           0.00609 &             0.00000 &             6.47786 &           0.00000 \\
        HOMO$^\mathrm{D}$   &           0.01300 &             0.00000 &             0.00000 &           5.76135 \\
    \end{tabular}
    \end{ruledtabular}
\end{table}

\begin{table}[ht]
    \caption{The electronic Hamiltonian elements for hole transfer of pair 2. (Unit: eV)}
    \begin{ruledtabular}
    \begin{tabular}{ccccc}
                            & HOMO$^\mathrm{A}$ & HOMO-2$^\mathrm{D}$ & HOMO-1$^\mathrm{D}$ & HOMO$^\mathrm{D}$ \\
        \hline
        HOMO$^\mathrm{A}$   &            6.56662 &            0.00825 &            -0.00313 &     0.00879 \\
        HOMO-2$^\mathrm{D}$ &            0.00825 &            7.43016 &             0.00000 &     0.00000 \\
        HOMO-1$^\mathrm{D}$ &           -0.00313 &            0.00000 &             6.51022 &     0.00000 \\
        HOMO$^\mathrm{D}$   &            0.00879 &            0.00000 &             0.00000 &     5.80322 \\
    \end{tabular}
    \end{ruledtabular}
\end{table}

\begin{table}[ht]
    \caption{The electronic Hamiltonian elements for hole transfer of pair 3. (Unit: eV)}
    \begin{ruledtabular}
    \begin{tabular}{ccccc}
                            & HOMO$^\mathrm{A}$ & HOMO-2$^\mathrm{D}$ & HOMO-1$^\mathrm{D}$ & HOMO$^\mathrm{D}$ \\
        \hline
        HOMO$^\mathrm{A}$   &           6.44302 &             0.00225 &            -0.01486 &           0.01402 \\
        HOMO-2$^\mathrm{D}$ &           0.00225 &             7.44463 &             0.00000 &           0.00000 \\
        HOMO-1$^\mathrm{D}$ &          -0.01486 &             0.00000 &             6.47786 &           0.00000 \\
        HOMO$^\mathrm{D}$   &           0.01402 &             0.00000 &             0.00000 &           5.81043 \\
    \end{tabular}
    \end{ruledtabular}
\end{table}

\begin{table}[ht]
    \caption{The electronic Hamiltonian elements for hole transfer of pair 4. (Unit: eV)}
    \begin{ruledtabular}
    \begin{tabular}{ccccc}
                            & HOMO$^\mathrm{A}$ & HOMO-2$^\mathrm{D}$ & HOMO-1$^\mathrm{D}$ & HOMO$^\mathrm{D}$ \\
        \hline
        HOMO$^\mathrm{A}$   &           6.82734 &             0.00732 &             0.00665 &          -0.00758 \\
        HOMO-2$^\mathrm{D}$ &           0.00732 &             7.34664 &             0.00000 &           0.00000 \\
        HOMO-1$^\mathrm{D}$ &           0.00665 &             0.00000 &             6.40686 &           0.00000 \\
        HOMO$^\mathrm{D}$   &          -0.00758 &             0.00000 &             0.00000 &           5.72391 \\
    \end{tabular}
    \end{ruledtabular}
\end{table}

\clearpage
\newpage
\section{Validation of one-electron approximation}
\begin{figure}[ht]
    \includegraphics[scale=0.35]{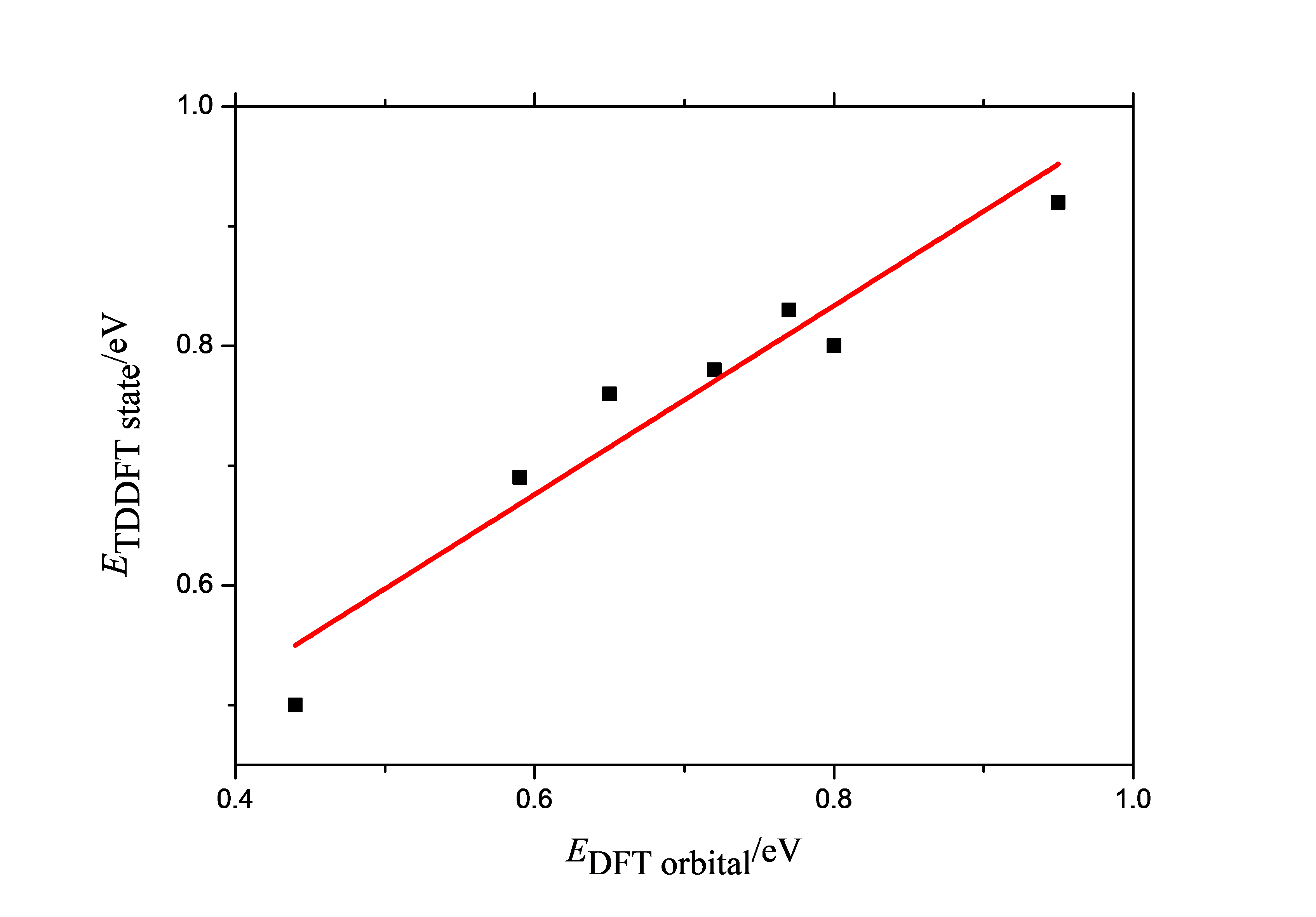}
    \caption{The energy comparison between one-electron approximation using DFT orbitals and multi-electron picture using TDDFT states. (We set the energies of HOMO$^\mathrm{D}$ in DFT and lowest hole state in TDDFT as zero)} \label{fig: 1}
\end{figure}

To validate the one-electron approximation, we perform TDDFT calculations of cationic D/A pairs to evaluate the electronic energies of multi-electron states. By examining the excitation characteristics of each state, we assign the low-lying states of TDDFT results as localised hole states. The results of state energies by TDDFT comparing to DFT orbital energies are shown in Fig.~\ref{fig: 1}.

\clearpage
\newpage
\section{Non-local spectral density}
\begin{figure}[ht]
    \includegraphics[scale=0.28]{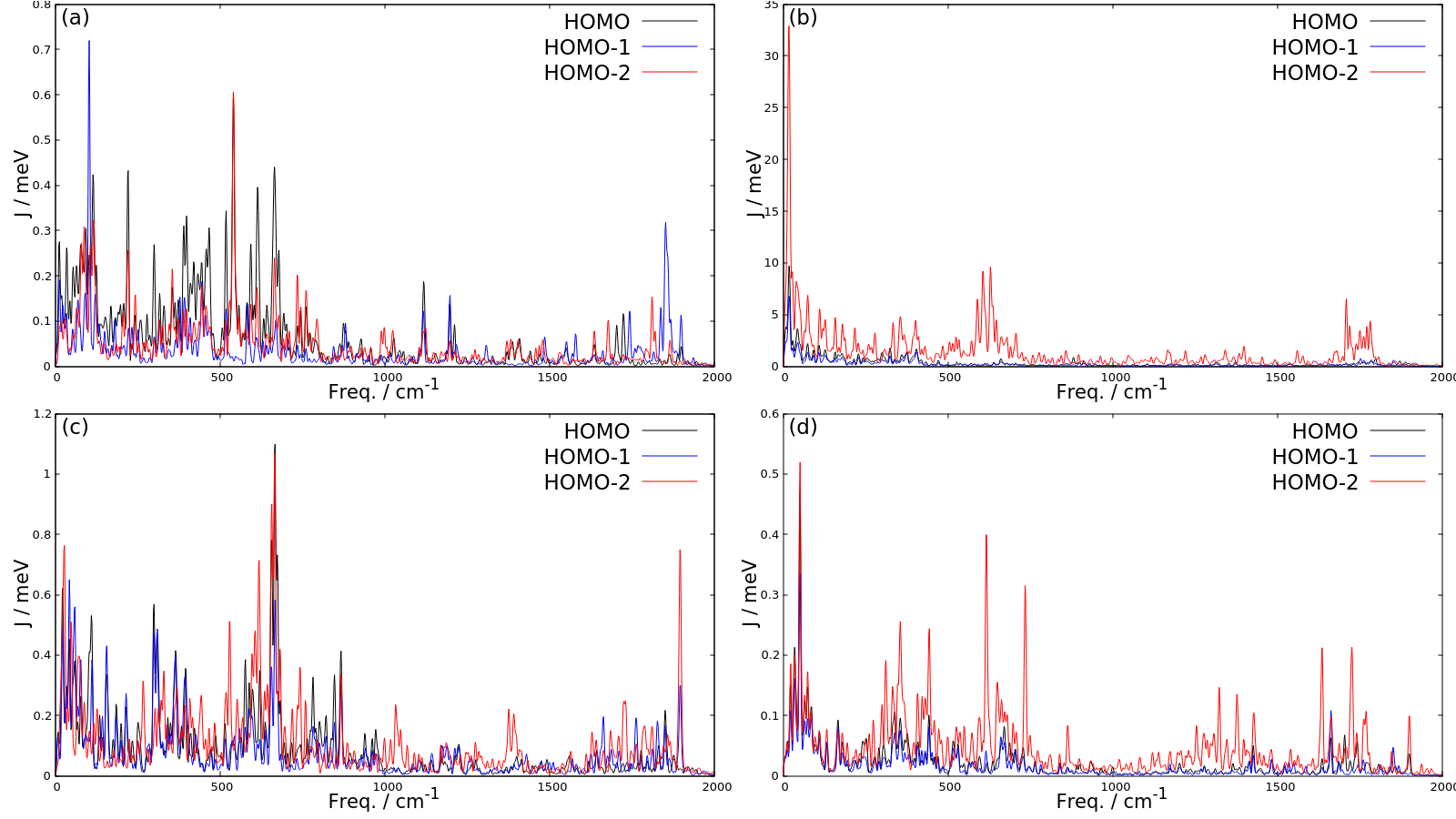}
    \caption{The spectral densities of the non-local el-vib couplings for electronic coupling between HOMO of acceptor and FMOs of donor for (a) pair 1, (b) pair 2, (c) pair 3 and (d) pair 4.} \label{fig: 2}
\end{figure}

\clearpage
\newpage
\section{Additional dynamics simulation}

\begin{figure}[ht]
    \includegraphics[scale=0.28]{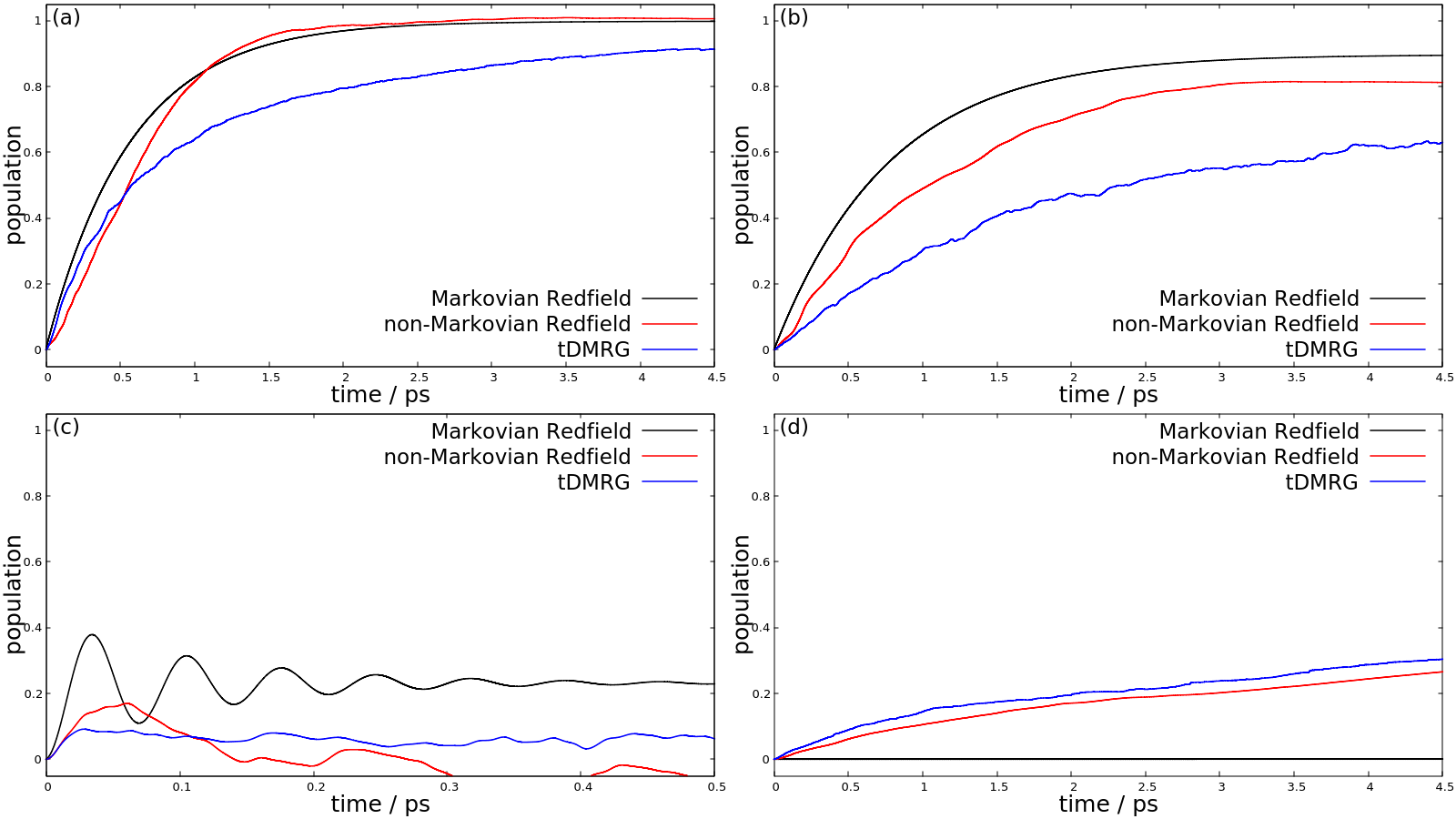}
    \caption{Comparison of time-evolution hole population at HOMO-1 of donor between the Markovian Redfield equation (black lines), non-Markovian Redfield (red lines) and tDMRG method (blue lines) for (a) pair 1, (b) pair 2, (c) pair 3 and (d) pair 4.} \label{fig: 3}
\end{figure}

For the model without non-local el-vib coupling terms, we apply original Redfield Eq.~11 in main text (i.e. non-Markovian Redfield) and time-dependent density matrix renormalization group (tDMRG) method, comparing with the results in Fig.~6 of main text.

For the non-Markovian Redfield equation, we just turn off the Markovian and secular approximation and use the same initial condition and model information (electronic Hamiltonian and local spectral densities) as Markovian Redfield equation we applied in main text. The results are shown in Fig.~\ref{fig: 3} using red lines.

For the tDMRG calculation, we adopte discretization of spectral density \cite{xie2019time} to get the local el-vib couplings,
\begin{equation}
g_{i}^{I}=\sqrt{\int_{\omega_I - \frac{1}{2}\Delta\omega}^{\omega_I + \frac{1}{2}\Delta\omega}\frac{2}{\pi}J_i(\omega)\mathrm{d}\omega}
\end{equation}
with a discrete series of bath frequencies $\{\omega_I\}$ with $\omega_{I+1}=\omega_I+\Delta\omega$ in the given region (0-0.25 eV) and we set $\Delta\omega = 0.01$ eV. i.e. 25 modes are coupled with each electronic state.

To include temperature in tDMRG calculation, we simply approximate the initial state as $\vert\mathrm{HOMO^A}\rangle\bigotimes_I\vert\langle\bar{n}_I\rangle\rangle$, with $\langle\bar{n}_I\rangle$ presents the rounding of average occupation number $\bar{n}_I$ of mode $I$. And $\bar{n}_I=(\exp\frac{\hbar\omega_I}{k_\mathrm{B}T} - 1)^{-1}$ with temperature $T=300$ K. The results are shown as blue lines in Fig.~\ref{fig: 3}.

According to the comparison in Fig.~\ref{fig: 3}, We could find that the Redfield calculations overestimate the transfer process for pair 1 and 2 but still give qualitatively similar pictures of deep hole transfer, while the Redfield theory with the secular and Markovian approximations gives an unphysical result for pair 4. For pair 3, the results are more complicated (maybe because of the large electron-electron and electron-vibration couplings corresponding to HOMO-1 of donor and the endothermal behaviour), the Markovian Redfield again overestimates the transfer while non-Markovian Redfield is crashed due to its non-positivity feature.

\bibliography{ref}